\documentclass[12pt]{article}
\usepackage{graphicx,epsfig,color,latexsym}
\newcommand{\bc}{\begin{center}}
\newcommand{\ec}{\end{center}}
\newcommand{\be}{\begin{equation}}
\newcommand{\ee}{\end{equation}}
\newcommand{\bea}{\begin{eqnarray}}
\newcommand{\eea}{\end{eqnarray}}
\newcommand{\ba}{\begin{array}}
\newcommand{\ea}{\end{array}}
\newcommand{\lb}{\label}
\newcommand{\rf}{\ref}
\newcommand{\bfg}{\begin{figure}[htbp]}
\newcommand{\efg}{\end{figure}}

\newcommand{\pr}{Phys. Rev. }
\newcommand{\np}{Nucl. Phys. }

\newcommand{\prp}{Phys. Rep. }

\newcommand{\pl}{Phys. Lett. }

\newcommand{\rmp}{Rev. Mod. Phys. }
\newcommand{\nc}{Nuovo Cimento }

\begin{document}

\begin{flushright}
IPNO-DR-08-04
\end{flushright}
\vspace{0.5 cm}
\bc

{\Large\textbf{Integral equation for gauge invariant\protect \\ 
\vspace{0.25 cm}
quark Green's function\footnote{Talk given at the Joint Meeting
Heidelberg-Li\`ege-Paris-Wroclaw, Spa, 6-8 March 2008.}}}\\
\vspace{1 cm}
H. Sazdjian\\
\textit{Institut de Physique Nucl\'eaire, CNRS/IN2P3,\\
Universit\'e Paris-Sud 11, F-91406 Orsay, France\\
\footnotesize{E-mail: sazdjian@ipno.in2p3.fr}}
\ec
\par
\vspace{2 cm}

\bc
{\large Abstract} 
\ec
We consider gauge invariant quark two-point Green's functions 
in which the gluonic phase factor follows a skew-polygonal line.
Using a particular representation for the quark propagator in
the presence of an external gluon field, functional relations 
between Green's functions with different numbers of segments 
of the polygonal lines are established. An integral equation 
is obtained for the Green's function having a phase factor 
along a single straight line. The related kernels involve Wilson 
loops with skew-polygonal contours and with functional derivatives 
along the sides of the contours. 
\par
\vspace{0.5 cm}
\par

\newpage

The objective of this work is to investigate the possibilities of
deriving integral or integrodifferential equations for the two-point
gauge invariant quark Green's function (GIQGF). Gauge invariant objects 
seem to have a firmer basis to describe non-pertur\-ba\-tive properties of 
QCD, as compared to gauge variant ones \cite{m1,nm}. A typical example of 
those is the Wilson loop \cite{w}, which allows the formulation of 
the confinement property of quarks with a relatively simple criterion 
\cite{bw,k}. 
\par
Gauge invariant Green's functions involve path-ordered gluon field
phase factors. Here, we concentrate on two-point quark
Green's functions, in which the gluonic phase factor follows
in general a skew-polygonal line.
\par
The starting point of our investigation is a particular representation 
for the quark propagator in the presence of an external gluon field, 
where it is expressed as a series of terms involving phase factors 
along successive straight lines. Then the corresponding quantized 
Green's function becomes expressed in terms of Wilson loops having 
skew-polygonal contours \cite{hs}.
\par  
We begin by introducing definitions and conventions.
We consider a path-ordered gluon field phase factor along a line 
$C_{yx}$ joining a point $x$ to a point $y$, with an 
orientation defined from $x$ to $y$:
\be \lb{e1}
U(C_{yx};y,x)\equiv U(y,x)=Pe^{{\displaystyle -ig\int_x^y 
dz^{\mu}A_{\mu}(z)}}.
\ee
Parametrizing the line $C$ with a parameter $\lambda$,
$C=\{x(\lambda)\}$, $0\leq \lambda\leq 1$, such that $x(0)=x$ and 
$x(1)=y$, a variation of $C$ induces the following
variation of $U$ [$U(x(\lambda),x(\lambda'))\equiv U(\lambda,\lambda')$,
$A(x(\lambda))\equiv A(\lambda)$] \cite{m1}:
\bea \lb{e2}
\delta U(1,0)&=&-ig\delta x^{\alpha}(1)A_{\alpha}(1)U(1,0)
+igU(1,0)A_{\alpha}(0)\delta x^{\alpha}(0)\nonumber \\
& &+ig\int_0^1d\lambda U(1,\lambda)x^{\prime \beta}(\lambda)F_{\beta\alpha}
(\lambda)\delta x^{\alpha}(\lambda)U(\lambda,0),
\eea
where $x'=\frac{\partial x}{\partial \lambda}$ and $F$ is the
field strength, 
$F_{\mu\nu}=\partial_{\mu}A_{\nu}-\partial_{\nu}A_{\mu}
+ig[A_{\mu},A_{\nu}]$.
\par
For paths defined along rigid lines, the variations inside the integral
are related, with appropriate weight factors, to those of the end points.
Considering now a rigid straight line between $x$ and $y$, 
an ordinary derivation at the end points yields:
\be \lb{e3}
\frac{\partial U(y,x)}{\partial y^{\alpha}}=-igA_{\alpha}(y)U(y,x)
+ig(y-x)^{\beta}\int_0^1d\lambda\,\lambda\, U(1,\lambda)F_{\beta\alpha}
(\lambda)U(\lambda,0),
\ee
\be \lb{e4}
\frac{\partial U(y,x)}{\partial x^{\alpha}}=+igU(y,x)A_{\alpha}(x)
+ig(y-x)^{\beta}\int_0^1d\lambda\,(1-\lambda)\, U(1,\lambda)F_{\beta\alpha}
(\lambda)U(\lambda,0).
\ee
\par
We adopt the following conventions to represent the contributions of 
the internal parts of the integrals:
\bea
\lb{e5}
& &\frac{\bar\delta U(y,x)}{\bar\delta y^{\alpha +}}\equiv
ig(y-x)^{\beta}\int_0^1d\lambda\,\lambda\, U(1,\lambda)F_{\beta\alpha}
(\lambda)U(\lambda,0), \\
\lb{e6}
& &\frac{\bar\delta U(y,x)}{\bar\delta x^{\alpha -}}\equiv
ig(y-x)^{\beta}\int_0^1d\lambda\,(1-\lambda)\, U(1,\lambda)F_{\beta\alpha}
(\lambda)U(\lambda,0).
\eea
\par
The Wilson loop, denoted $\Phi(C)$, is defined as the trace, in color 
space, of the phase factor (\rf{e1}) along a closed contour $C$:
\be \lb{e7}
\Phi(C)=\frac{1}{N_c}\mathrm{tr}Pe^{{\displaystyle -ig\oint_C
dx^{\mu}A_{\mu}(x)}}.
\ee
Its vacuum expectation value is denoted $W(C)$:
\be \lb{e8}
W(C)=\langle\Phi(C)\rangle,
\ee
the averaging being defined in the path integral formalism. The 
properties of the Wilson loop were studied in a long series of
papers \cite{p,mm1,mm2,mgd,mk,m2,g,kkk,br,dv,bnsg,bmpbbp,bcpbmp,sdks}.
We adopt the following functional representation for $W(C)$:
\be \lb{e9}
W(C)=e^{{\displaystyle F(C)}}.
\ee
In perturbation theory, $F(C)$ is given by the sum of all 
connected diagrams, the connection being defined with respect to the 
contour $C$ \cite{dv}. For large contours and large $N_c$, $F(C)$ is 
proportional to the area of the minimal surface with contour $C$ 
\cite{mm1,mm2}.
\par
If the contour $C$ is a skew-polygon $C_n$ with $n$ sides and 
$n$ successive marked points $x_1$, $x_2$, $\ldots$, $x_n$ at the 
cusps, then we write:
\be \lb{e10}
W(x_n,x_{n-1},\ldots,x_1)=W_n=
e^{{\displaystyle F_n(x_n,x_{n-1},\ldots,x_1)}}
=e^{{\displaystyle F_n}}.
\ee
\par
The two-point GIQGF, with a phase factor along a line $C$, is defined as
\be \lb{e11}
S_{\alpha\beta}^{}(x,x';C_{x'x})=-\frac{1}{N_c}\,\langle\overline 
\psi_{\beta}^{}(x')\,U(C_{x'x};x',x)\,\psi_{\alpha}^{}(x)\rangle.
\ee
(The color indices are implicitly summed.)
For skew-polygonal lines with $n$ sides and $n-1$ junction points 
$y_1$, $y_2$, $\ldots$,$y_{n-1}$ between the segments, we define:
\be \lb{e12}
S_{(n)}(x,x';y_{n-1},\dots,y_1)=-\frac{1}{N_c}\,\langle\overline \psi(x')
U(x',y_{n-1})U(y_{n-1},y_{n-2})\ldots U(y_1,x)\psi(x)\rangle.
\ee
For one straight line, one has:
\be \lb{e13}  
S_{(1)}(x,x')\equiv S(x,x')=-\frac{1}{N_c}\,\langle\overline \psi(x')
\,U(x',x)\,\psi(x)\rangle.
\ee
( The index 1 will generally be omitted from that function.)
\par
We shall adopt a two-step quantization method. One first integrates
with respect to the quark fields. This produces in various terms the 
quark propagator in the presence of the gluon field. Then one
integrates with respect to the gluon field through Wilson loops.
\par
To make Wilson loops appear, one needs an appropriate 
representation for the quark propagator in extenal field. 
We use the following representation which involves phase factors
along straight lines together with the full quark Green's function
$S_{(1)}\equiv S$ \cite{hs,js}:
\be \lb{e14}
S(x,x';A)=S(x,x')U(x,x')+
\Big(S(x,y)\frac{\bar\delta U(x,y)}{\bar\delta y^{\alpha -}}
+\frac{\bar\delta S(x,y)}{\bar\delta y^{\alpha +}}U(x,y)\Big)
\gamma^{\alpha}S(y,x';A),
\ee
where $S(x,x';A)$ is the quark propagator in the presence of the
external gluon field $A$. (Integrations on intermediate points are 
implicit.) This equation yields an expansion of $S(x,x';A)$ in terms
of the GIQGF $S$ and explicit phase factors along straight lines. 
The above representation is a generalization of that used for heavy 
quark propagators starting from the static case \cite{ef}.
\par
Systematic use of Eq. (\rf{e14}) leads to functional relations 
between various GIQGFs.
Consider for this the Green's function $S_{(n)}$ [Eq. (\rf{e12})]. 
Integrate with respect to the quark fields; one obtains:
\be \lb{e15}
S_{(n)}(x,x';y_{n-1},\ldots,y_1)=\frac{1}{N_c}\,\langle U(x',y_{n-1})
U(y_{n-1},y_{n-2})\cdots U(y_1,x)S(x,x';A)\rangle.
\ee
Use of the expansion (\rf{e14}) for $S(A)$ gives:
\bea \lb{e16}
& &S_{(n)}(x,x';y_{n-1},\ldots,y_1)=
S(x,x')\,e^{{\displaystyle F_{n+1}(x',y_{n-1},\ldots,y_1,x)}}
\nonumber \\
& &\ \ \ \ +\Big(\frac{\bar\delta S(x,y_n)}{\bar\delta y_n^{\alpha +}}
+S(x,y_n)\frac{\bar\delta }{\bar\delta y_n^{\alpha -}}\Big)
\,\gamma^{\alpha}\,S_{(n+1)}(y_n,x';y_{n-1},\ldots,y_1,x).
\eea
Notice the appearance of the Wilson loop average along the 
skew-polygonal contour with $(n+1)$ sides.
Repeating the same operation for $S_{(n+1)}$ and so forth, one
can in principle express any $S_{(n)}$ in terms of the lowest-order
Green's function $S$ and Wilson loop averages and their derivatives
along the corresponding contours.
\par
Next, we use the equation of motion of the quark fields. This yields
for $S_{(n)}$:
\bea \lb{e17}
& &(i\gamma.\partial_{(x)}-m)S_{(n)}(x,x';y_{n-1},\ldots,y_1)=
i\delta^4(x-x')e^{{\displaystyle F_{n}(x,y_{n-1},\ldots,y_1)}}
\nonumber \\
& &\ \ \ \ \ \ \ \ 
+i\gamma^{\mu}\frac{\bar\delta S_{(n)}(x,x';y_{n-1},\ldots,y_1)}
{\bar\delta x^{\mu -}}.
\eea
\par
We observe that the right-hand side contains, as an unknown, the
functional derivative along the rigid segment $xy_1$ of $S_{(n)}$.
The task of evaluating that quantity is, however, facilitated by
the functional relations (\rf{e16}), which relate two 
successive Green's functions with increasing index. They allow the 
evaluation of the rigid path derivative of a Green's function in terms 
of a similar derivative of a Wilson loop average and the derivative of 
a Green's function with a higher index. Systematic repetition of this 
procedure allows us to express the derivative of a Green's 
function in terms of a series of Green's functions whose coefficients 
are functional derivatives of Wilson loop averages. One thus obtains
chains of coupled integral (or integrodifferential) equations between 
the various Green's functions. At the end, each Green's function
$S_{(n)}$ can be expressed, at leading order of an expansion, by
means of the functional relation (\rf{e16}), in terms of the 
lowest-order Green's function $S$. Thus an equation where solely
the Green's function $S$ would appear becomes reachable.
\par
In the present work we are mainly interested by the simplest Green's 
function $S$; the general structure of the derivative 
$\bar\delta S/\bar\delta x^{\mu -}$ is:
\bea \lb{e18}
\frac{\bar\delta S(x,x')}{\bar\delta x^{\mu -}}&=&K_{1\mu -}(x',x)\,
S(x,x')+K_{2\mu -}(x',x,y_1)\,S_{(2)}(y_1,x';x)\nonumber \\
& &\ +\sum_{n=3}^{\infty}K_{n\mu -}(x',x,y_1,\ldots,y_{n-1})\,
S_{(n)}(y_{n-1},x';x,y_1,\ldots,y_{n-2}).
\eea
The kernels $K_n$ ($n=1,\ldots,\infty$) are composed of Wilson
loop averages along skew-polygonal contours with $(n+1)$ sides,
of their derivatives along these sides and of $(n-1)$ quark Green's
functions $S$ and their derivatives.
The total number of derivatives contained
in $K_n$ is $n$. Each segment of the contour is submitted at most to
one derivation. The calculation of the expression of the kernel $K_n$ 
requires solely consideration of terms of order lower or equal to $n$.
\par
The explicit expression of $\bar\delta S/\bar\delta x^{\mu -}$ up to 
the third-order of its expansion is:
\bea \lb{e19} 
& &\frac{\bar\delta S(x,x')}{\bar\delta x^{\mu -}}=
\frac{\bar\delta F_{2}(x',x)}{\bar\delta x^{\mu -}}\,S(x,x')
-\frac{\bar\delta^2 F_3(x',x,y_1)}{\bar\delta x^{\mu -}
\bar\delta y_1^{\alpha_1 +}}\,S(x,y_1)\,\gamma^{\alpha_1}\,
S_{(2)}(y_1,x';x)
\nonumber \\
& &\ \ \ \ +\frac{\bar\delta^3 F_4(x',x,y_1,y_2)}{\bar\delta x^{\mu -}
\bar\delta y_1^{\alpha_1 +}\bar\delta y_2^{\alpha_2 +}}\,S(x,y_1)\,
\gamma^{\alpha_1}\,S(y_1,y_2)\,\gamma^{\alpha_2}\,S_{(3)}(y_2,x';x,y_1)
\nonumber\\
& &\ \ \ \ +\frac{\bar\delta^2 F_4}{\bar\delta x^{\mu -}
\bar\delta y_2^{\alpha_2 +}}\,S(x,y_1)\,\gamma^{\alpha_1}\, 
\Big(\frac{\bar\delta S(y_1,y_2)}{\bar\delta y_1^{\alpha_1 -}}
+S(y_1,y_2)\frac{\bar\delta F_4}{\bar\delta y_1^{\alpha_1 +}}\Big)\,
\gamma^{\alpha_2}\,S_{(3)}(y_2,x';x,y_1)\nonumber \\
& &\ \ \ \ +\cdots\ \ .
\eea
The contents of the kernels $K_{n}$ can also be classified according
to well-known structures, such as completely connected, crossed, 
nested, etc. 
\par
The integral form of the equation of motion (\rf{e17}) for $n=1$ is:
\be \lb{e20}
S(x,x')=S_0(x,x')+\int d^4x''\,S_0(x,x'')\,\gamma^{\mu}\,
\frac{\bar\delta S(x'',x')}{\bar\delta x^{''\mu -}},
\ee
in which one has to inject the expression of 
$\bar\delta S/\bar\delta x^-$ resulting from Eq. (\rf{e18}).
($S_0$ is the free quark propagator.)
\par 
At short distances, governed by perturbation theory, each 
derivation introduces a new power of the coupling constant and 
therefore the dominant terms in the expansion are the lowest-order 
ones. At large distances, Wilson loops are saturated by the minimal 
surfaces having as supports the contours 
\cite{mm1,mm2,bmpbbp,bcpbmp,sdks,js}. Here 
also, the dominant contributions come from the lowest-order derivative 
terms. Therefore, the expansion in Eq. (\rf{e20}) can be considered in 
general as perturbative whatever the distances are, provided that 
for each type of region the appropriate expressions are used for the 
Wilson loops. The first term of the expansion, represented by a single 
derivative, is null for symmetry reasons (the derivative 
$\bar\delta F_{2}(x',x)/\bar\delta x^-$ being orthogonal to $xx'$).
Hence the non-zero dominant term of the expansion is the second-order
derivative one. Furthermore, the various Green's functions $S_{(n)}$
are themselves dominated by their lowest-order expression of Eq.
(\rf{e17}), involving only $S$ and a Wilson loop. In that approximation,
$\bar\delta S(x,x')/\bar\delta x^-$ takes the form
\be \lb{e21}
\frac{\bar\delta S(x,x')}{\bar\delta x^{\mu -}}\simeq
-\int d^4y_1\,\frac{\bar\delta^2 F_3(x',x,y_1)}{\bar\delta x^{\mu -}
\bar\delta y_1^{\alpha_1 +}}\,
e^{{\displaystyle F_3(x',x,y_1)}}\,S(x,y_1)\,\gamma^{\alpha_1}\,
S(y_1,x').
\ee
Thus, the dominant part of the integral equation (\rf{e20}) to be 
solved reduces to a closed form relative to the full Green's 
function $S$.
\par
The approximate form (\rf{e21}) can be used for a first resolution
of the integral equation (\rf{e20}).
\par

\vspace{0.5 cm}
\noindent
\textbf{Acknowledgements}
\par
This work was supported in part by the EU network FLAVIANET under 
Contract No. MRTN-CT-2006-035482.
\par

\end{document}